\def\spose#1{\hbox to 0pt{#1\hss}}
\def\lta{\mathrel{\spose{\lower 3pt\hbox{$\mathchar"218$}}
     \raise 2.0pt\hbox{$\mathchar"13C$}}}
\def\gta{\mathrel{\spose{\lower 3pt\hbox{$\mathchar"218$}}
     \raise 2.0pt\hbox{$\mathchar"13E$}}}
\def\ge{\mathrel{\spose{\lower 3pt\hbox{$-$}}
     \raise 2.0pt\hbox{$\mathchar"13E$}}}
\def\le{\mathrel{\spose{\lower 3pt\hbox{$-$}}
     \raise 2.0pt\hbox{$\mathchar"13C$}}}

\documentclass[aps,superscriptaddress]{revtex4}

\usepackage{graphics}

\begin{document}

\bibliographystyle{apsrev}

\title{Journey to the edge of time: The GREAT mission.}
\author{Neil J. Cornish}
\affiliation{Department of Physics, Montana State University, Bozeman, MT 59717}
\author{David N. Spergel}
\affiliation{Department of Astrophysical Sciences, Peyton Hall,
Princeton University, Princeton, NJ 08544-1001, USA}
\author{Charles L. Bennett}
\affiliation{Code 685, Laboratory for Astronomy \& Solar Physics, Goddard Space Flight Center,
Greenbelt, MD 20771}

\begin{abstract}
We are surrounded by radiation that originated from the big bang. It has traveled to us
from the farthest reaches of the Universe, carrying with it an unaltered record of the
beginning of time and space. The radiation is in the form of gravitational waves -
propagating ripples in the curvature of spacetime. We describe a mission to detect these
Gravitational Echos Across Time (GREAT) that would open up a new window on the very
early universe. By studying the gravitational echoes of the big bang we will gain insight into
the fundamental structure of matter, gravity, and how the Universe formed.\\

Submitted to NASA's 2003 SEU Roadmap Team as a ``Vision'' mission. Vision missions
typically require technologies that are not yet developed - this is certainly
true of GREAT.
\end{abstract}

\maketitle

\section{Introduction}
The Universe today is filled with fossil relics of the big bang. Studies of one of these
relics - the Cosmic Microwave Background (CMB) - have revolutionized our understanding
of the early Universe. But microwaves and other forms of
electromagnetic radiation are unable to penetrate the hot plasma that filled
the Universe until 300,000 years after the big bang. To see beyond this recombination barrier
we need to study another fossil relic of the big bang - the Cosmic Gravitational
wave Background (CGB). These gravitational waves were produced when the Universe first
formed, and their extremely weak coupling to matter allowed them to propagate
freely through the dense plasma that filled the early Universe. In most models of the
early Universe the CGB has a quantum mechanical origin, making its detection a probe of quantum
gravity.

It is not easy to detect waves that can pass virtually unattenuated through the
densest matter in the Universe, but the task is not impossible.
A GREAT detector consisting of two independent gravitational wave interferometers that
are enhanced versions of the LISA\cite{lisa} instrument (Laser Interferometer Space Antenna) would
have the sensitivity required to detect the Cosmic Gravitational wave Background predicted
by standard inflation models. Indeed, the main challenge is not building a detector with
sufficient sensitivity, but rather, picking out the CGB from the multitude of astrophysical
sources of gravitational waves. The final design of the GREAT mission will depend on
lessons learned from the LISA mission. LISA will serve as a stepping stone in the development
of the advanced optical and drag free systems required by GREAT, while at the same time
providing crucial data about the astrophysical foregrounds that GREAT will have to contend
with.

\section{Why go to space?}

The gravitational wave spectrum is broadly divided into high and low frequencies, with
one Hertz as the dividing line. Seismic, atmospheric and other environmental disturbances
limit ground based gravitational wave detectors to the high frequency portion of the
spectrum above 10 Hertz. In contrast, space based detectors can operate at practically any
frequency. 

Standard inflation theories predict that the amplitude, $h$, of the relic gravitational waves
increases at low frequencies such that $h \sim f^{-\alpha}$, where $f$ is the frequency
and $\alpha \approx 1.5$. According to this prediction, the gravitational wave background
in the frequency range accessible from space will have an
amplitude thousands to trillions of times greater than in the frequency range
accessible on Earth.

\section{Science Potential}

The CGB contains a wealth of information about the very early Universe. The amplitude
and energy spectrum of the CGB provides a sensitive probe of the physical processes at
work during the formation of the Universe. The gravitational waves produced during the
big bang are expect to have a quantum origin, making the CGB a direct experimental
test of quantum gravity. The leading theory of the early universe, Inflation, predicts
that gravitational waves will be produced as the result of parametric amplification of
quantum fluctuations in the gravitational field. More speculative string theory
models of the early universe give a range of predictions for the CGB spectrum that
differ significantly from standard Inflation models. A direct measurement of the CGB
power spectrum would help to pin down the correct theory of quantum cosmology.

To date, observations have told us very little about the CGB power spectrum. The existing
experimental bounds are shown in Fig. 1, along with the regions of the spectrum that
will be probed by various gravitational wave detectors. Also shown is the prediction
from standard Inflation\cite{turner} with a tensor-scalar ratio of $0.01$. The spectrum
is expressed in terms of $\Omega_{gw}(f)$ - the energy density of gravitational waves, in units of
the critical density, per logarithmic frequency interval. Neither the ground based LIGO
(Laser Interferometric Gravitational Observatory) detectors, nor the space based
LISA detector, will have the sensitivity required to detect the spectrum predicted by
Inflation. To do so will require the GREAT detectors described in the next section.

If we  can achieve this sensitivity, then the GREAT mission has the potential to
probe physics at the Planck scale, test the inflationary paradigm and to detect
the creation of matter at the electroweak phase transition.  Next, we highlight
four of the potential important science questions probed by GREAT.

\subsection{Inflationary models}

The combination of a CMB polarization mission, such as CMBPOL, that
measures gravity modes on the horizon size\cite{pet} and the
GREAT mission will enable us to probe
the shape of the inflaton potential.  Since the amplitude of
the gravity wave signal depends on the Hubble parameter when the mode
leaves the horizon, we will be able to directly probe inflationary
physics.
The lever arm provided by the two measurements will be very powerful.
We will be probing the physics of inflation at two different scales
and will be able to perhaps reconstruct the inflationary potential
\cite{turner,hut}

\subsection{Probing the Planck Scale}

String theorists suggest that Planck scale physics can produce
a dramatic signature in the gravity wave background.  Eather et al.\cite{eat}
argue that the breakdown of locality in string theory can produce
significant deviations in the amplitude and shape of the gravity
wave spectrum.
Kaloper et al.\cite{kal} show that even if string theory
does not violate locality it will modify the Lagrangian by
adding non-minimal
coupling terms of order $p^2/M^2$ where $M$ is the string mass scale.
They argue  that string theories give rise to correction terms
of order $10^{-1} - 10^{-7}$ rather than the 10$^{-11}$ amplitude
terms that arise in inflationary models.  
These terms alter the relationship between
the tensor mode slope and the ratio between the amplitude of the tensor
mode signal and the amplitude of the scalar mode signal.
Since these correction terms are scale dependent, they will have
different effects on gravity waves at the two different scales.
This scale dependance appears in many versions of string cosmology
\cite{kal,all,men,inf}.

\subsection{TeV-PeV scale physics: the origin of matter and forces}

After inflation, the universe begins to cool and slowly expand.  GREAT
is probing the physics of the universe when its temperature was in the 
GeV - PeV range.  During this period, the universe underwent a series
of phase transitions.  These phase transitions, if they are violent
enough, will produce a detectable signal in the gravity wave background\cite{kos}.
In standard physics models, the universe went through two important
phase transitions in the GREAT frequency range: the QCD phase transition
during which quarks where bound into protons and neutrons and the
electroweak phase transition that transformed the basic forces into
the form that we observe them today.  While the gravity wave
signal in minimal electroweak models is too small for detection 
($h \sim 2 \times 10^{-27}$ at a characteristic frequency of $4\times
10^{-3}$ Hz), gravity
waves from more strongly first-order phase transitions, including the
electroweak transition in nonminimal models, have better prospects for
detection\cite{kam}.  Electroweak baryogensis
requires non-minimal models\cite{trod}, so that
a gravity wave background detection could be a direct signal of the
origin of matter itself!

In brane world models, the largest compact dimension can produce
an imprint on the gravity wave background.  In these models, the
fundamental scale of gravity is only 1 - 1000 TeV.  When the universe
cooled through this temperature, it can generate gravity waves
in the detectable range of LISA and GREAT\cite{ha,hb}

\subsection{Black hole Mergers: Science from the Foreground}

The greatest obstacle to detecting the CGB - the astrophysical foregrounds - can also be
viewed as a major science goal for the GREAT observatory. A very low frequency mission
would be able to detect every compact supermassive black hole binary in the Universe, while
an intermediate frequency mission could detect every compact Neutron star binary in the
Universe. The supermassive black hole binaries would provide insight into galactic
evolution and merger, while the Neutron star binaries would provide insight into the
endpoint of stellar evolution.

\section{Mission Design}

The GREAT detectors could potentially operate anywhere in the range $10^{-7} \rightarrow 10$ Hz.
Going below $10^{-7}$ Hz is not practical as observing a single cycle would take over a
year, while frequencies above 10 Hz are best handled by ground based detectors. The basic
design of the detectors would be the same as for LISA: laser interferometers monitoring the
separation of three or more free-floating proof masses that are shielded from
non-gravitational disturbances by the surrounding spacecraft. The GREAT mission will
likely differ from LISA in the number of interferometers, orbital configuration, baseline,
laser power, optics, and drag-free performance. The technologies required are well beyond
what are currently available, but there is no fundamanetal reason why the performance
levels can not be reached.

The final design will depend on the
frequency window that is chosen, and the choice of frequency window will depend on what
we learn about astrophysical foregrounds in the next decade. Based on our current
understanding of the foregrounds, there appear
to be two distinct choices for the GREAT observatory\cite{uv}.
The first is a very low frequency option\cite{cl}, operating in the range $10^{-7}
\rightarrow 10^{-5}$ Hz, that exploits the expected increase in the
amplitude of the CGB at low frequencies, and a tailing off in the amplitude of
astrophysical sources below $10^{-5}$ Hz. The second is an intermediate frequency option,
operating in the range $10^{-2} \rightarrow 1$ Hz, that exploits the transient nature
of the astrophysical
foregrounds at frequencies above $10^{-2}$ Hz. Each option has its own set of pros
and cons, and its own set of technical challenges. In either case, it is important that
the noise level in the observatory is well below the CGB signal so that
astrophysical foregrounds can be removed. 

\subsection{Very Low Frequency Mission}

The key technology for low frequency operation is the drag-free system that shields
the proof mass from non-gravitational disturbances. At the heart of the drag-free
system are the acceleration detectors. For a very low frequency mission to be viable,
accelerometer performance would need to improve by several orders of magnitude
beyond that envisioned for the LISA mission. It is unlikely that accelerometer improvements
alone will be enough to bridge the gap between LISA's low frequency performance and
the requirements of the GREAT observatory. The baseline of the interferometer will also
have to be increased, which will necessitate a change in the type of orbit used.
The LISA design employs a cartwheeling orbit with a five million kilometer baseline.
Changing to circular orbits with the spacecraft arrayed about the Sun leads to a
factor of 50 increase in the baseline while maintaining a one AU orbital radius.
The main drawback of this orbital configuration is a lack of directional sensitivity
compared to the LISA orbits. It may be necessary to increase the baseline further by
increasing the orbital radius to Martian or even Neptunian scale. This has the added
advantage of reducing thermal noise in the spacecraft, but at the cost of solar power
to run the spacecraft. Contrary to ones natural intuition, the larger baseline missions
do not require larger telescopes or more powerful lasers than the LISA mission.

To give a concrete example, a very low frequency GREAT observatory
with a $2.6 \times 10^8$ km baseline and accelerometers sensitive to accelerations of
$10^{-18}$ m s$^{-2}$ Hz$^{-1/2}$ would be able to detect a gravitational wave
background generated during the inflationary epoch at frequencies between
$10^{-6}$ and $10^{-5}$ Hz (Fig. 2).

Once the desired sensitivity has been reached, a method has to be found to distinguish
between instrument noise and the CGB, as both produce a stochastic signal in the
detector. One method is to simultaneously operate the observatory in what are know
as Sagnac and Micheleson modes. The Sagnac mode is fairly insensitive to low frequency
gravitational waves, making it a useful tool for monitoring the noise in the more
sensitive Michelson mode\cite{aet}. The advantage of this system is that it can be implemented
using a single interferometer. The other alternative is to fly two independent
interferometers and cross-correlate their output\cite{c}. The gravitational wave signal will
be common to both interferometers, while the detector noise will be independent.
Consequently, the signal to noise ratio grows as the square-root of the number of
cycles observed. The drawback of operating at low frequencies is that few complete cycles
can be observed in one year, making the cross correlation of two interferometers
inefficient.

\subsection{Intermediate Frequency Mission}

An intermediate frequency mission would require significant improvements in laser,
optical and drag-free technology. To accurately track the movements of the proof
masses requires a large number of photons. The more photons the better the measurement.
The position sensing improves as the square of the size of the telescopes used to
transmit and receive the laser signals, and as the square root of the laser power. 
Larger telescopes also reduce the error caused by laser pointing instability. The baseline
of an intermediate frequency mission has to be shorter than the LISA mission to
avoid losing sensitivity to waves whose wavelength fits inside the interferometer.

As a concrete example, a GREAT observatory comprising two
independent interferometers, each with $5\times 10^5$ Km baselines, 100 W lasers,
8 m telescopes, and accelerometers capable of detecting $10^{-17}$ m s$^{-2}$ Hz$^{-1/2}$,
would have the sensitivity required to detect gravitational waves
produced during an inflationary epoch of the early universe (Fig. 3).

The increased laser power and
large optics require a different spacecraft design from the LISA mission. The telescope
design could be based on Next Generation Space Telescope (NGST) technology. To have
two independent interferometers, a total of 8 spacecraft are required. They are arranged in
a square with two spacecraft at each corner. The spacecraft would have one large telescope
for sending and receiving signals along the interferometer arms, and a second smaller telescope
for exchanging phase information between the corner pair (Fig. 4).

A more radical design option for an intermediate frequency mission would be to shorten
the baseline further and use active postion control to operate the detector as
a power-recyled Fabry-Perot interferometer, as is done at the Earth based observatories.

\begin{figure}[ht]
\vspace{70mm}
\includegraphics{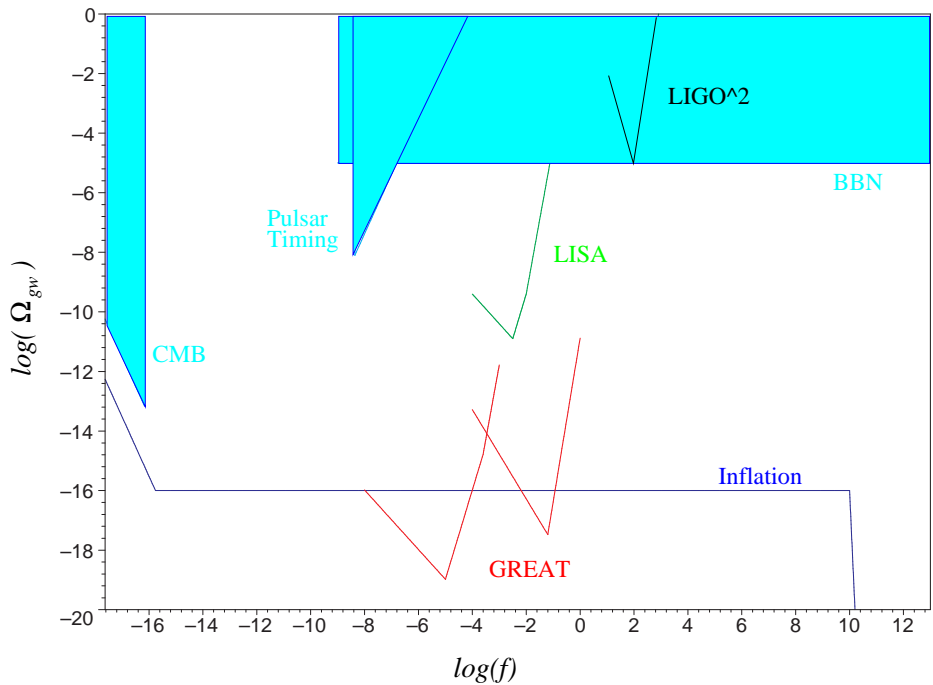}
\vspace{20mm}
\caption{Observational bounds on the CGB power spectrum, $\Omega_{gw}(f)$, are indicated by
the solid light blue boxes. The bounds are derived from the CMB anisotropy, Pulsar timing
and Big Bang Nucleosynthesis. The region of the spectrum that will be probed by
cross-correlating the first generation of LIGO detectors is indicated in black. The regions
of the spectrum that would be probed by LISA and the GREAT observatories are indicated
in green and red respectively. The blue curve is the spectrum predicted by Inflation with
a tensor-scale ratio of $0.01$.}
\end{figure}

\begin{figure}[ht]
\vspace{70mm}
\includegraphics{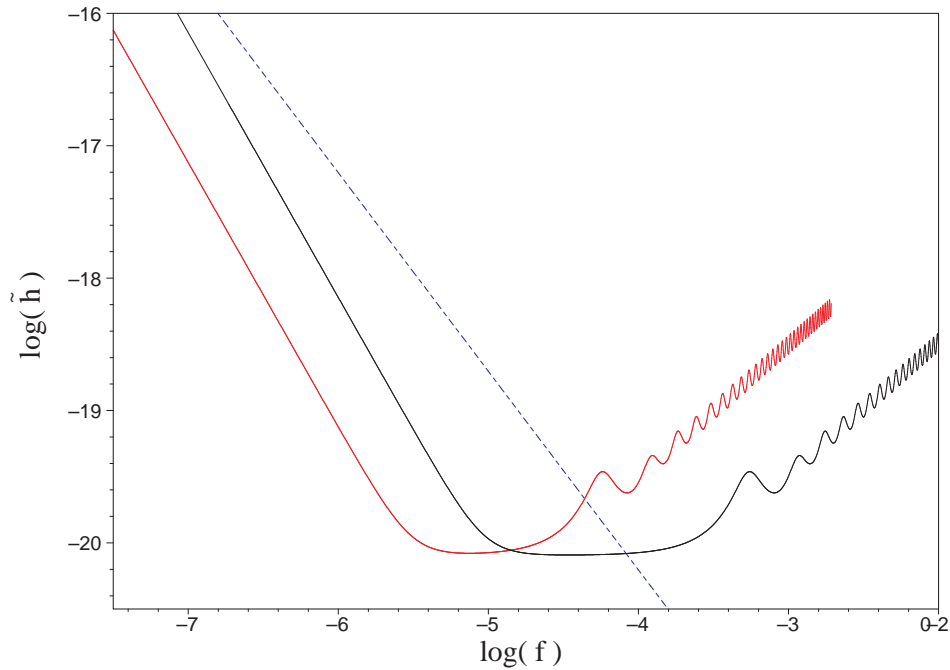}
\vspace{20mm}
\caption{The solid lines are sensitivity curves for two variants of a VLF mission.
The one to the right is described in the text, while the one on the left orbits
at 10 AU. The dotted line indicates the amplitude of the CGB in an inflationary
model with $\Omega_{gw}=10^{-16}$. The frequency, $f$ is measured in Hz, and the
strain spectral density $\tilde{h}$ is measured in Hz$^{-1/2}$.}
\end{figure}

\begin{figure}[ht]
\vspace{70mm}
\includegraphics{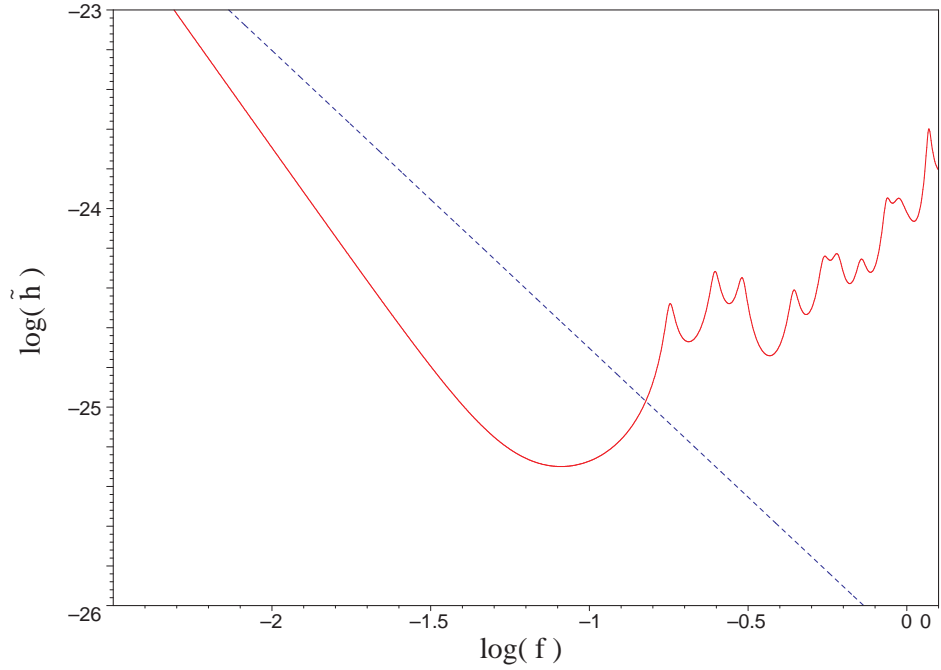}
\vspace{20mm}
\caption{The solid line shows the sensitivity that can be achieved by cross-correlating the
intermediate frequency interferometers described in the text. The correlation is
performed for one year with a frequency resolution of
$\log(f)=0.1$. The frequency, $f$ is measured in Hz, and the
strain spectral density $\tilde{h}$ is measured in Hz$^{-1/2}$.}
\end{figure}

\begin{figure}[ht]
\vspace{75mm}
\includegraphics{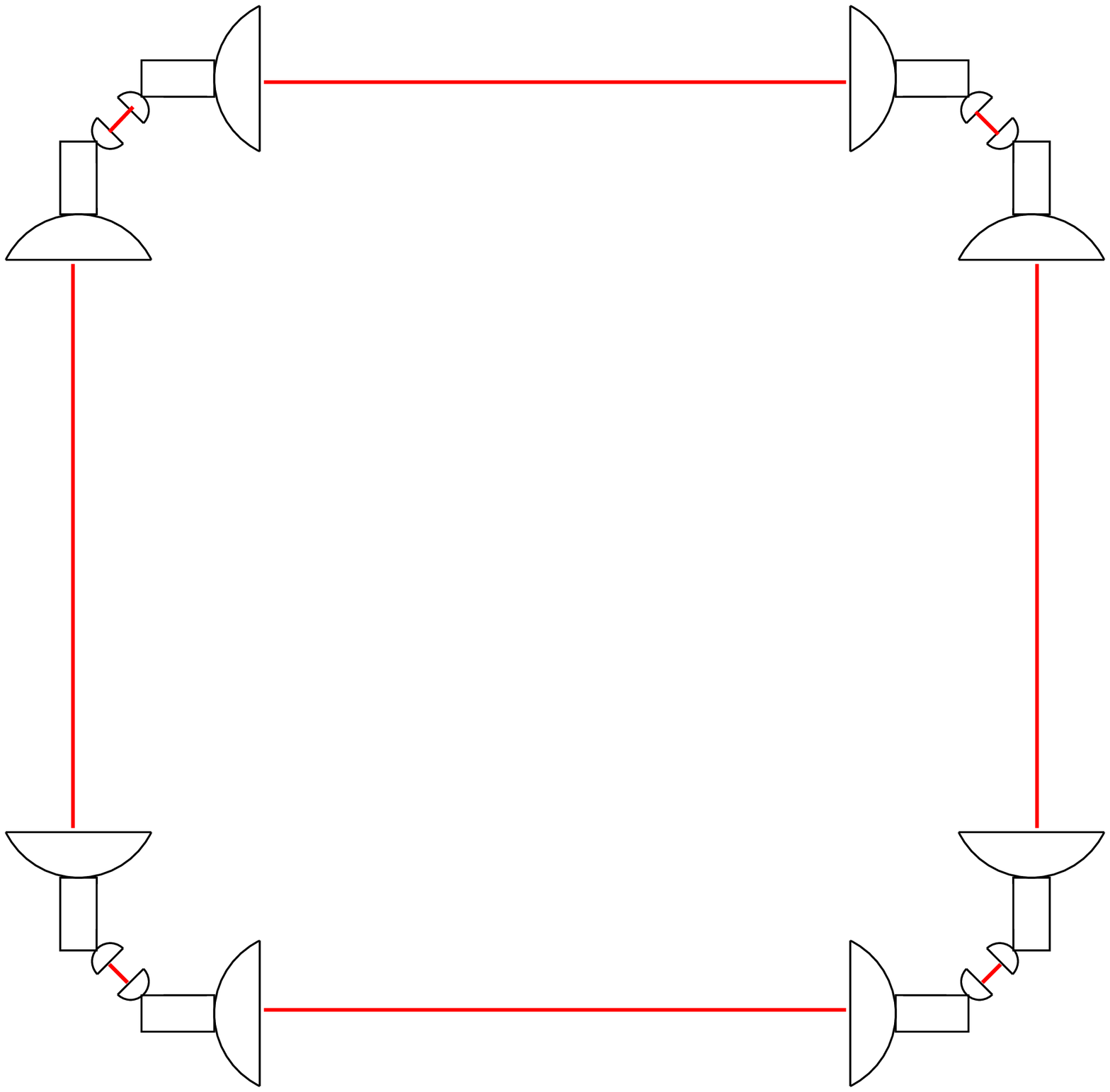}
\vspace{20mm}
\caption{The relative orbital configuration that gives two independent interferometers.}
\end{figure}


\begin{thebibliography}{99}
\bibitem{lisa} P. L. Bender {\it et al.}, {\it LISA Pre-Phase A Report} (1998).
\bibitem{turner} M. S. Turner,  Phys. Rev. D{\bf 55}, 435 (1997).
\bibitem{pet} J. B. Peterson, J. E. Carlstrom, E. S. Cheng, M. Kamionkowski,
A. E. Lange, M. Seiffert, D. N. Spergel, A. Stebbins, astro-ph/9907276.
\bibitem{hut} D. Huterer \& M.S. Turner, PRD, 62, 3503 (2000).
\bibitem{eat} R. Eather, B. Greene, W.H. Kinney, \& G. Shiu, hep-th/0110226 (2001).
\bibitem{kal} N. Kaloper, M. Kleban, A. Lawrence,  \& S. Shenker, hep-th/0201158(2002).
\bibitem{all} B. Allen \& R. Brustein, PRD, 55, 3260 (1997).
\bibitem{men} L. Mendes  \& A.R. Liddle, PRD, 60, 3508 (1999).
\bibitem{inf} M.P. Infante \& N. Sanchez, PRD, 61, 35151 (2000).
\bibitem{kos}  A. Kosowsky, M.S. Turner, \& R. Watkins, PRL, 69, 2026 (1992).
\bibitem{kam} M. Kamionkowski, A. Kosowsky \& M. Turner, PRD, 49, 2837 (1994).
\bibitem{trod} M. Trodden, RMP, 71, 1463 (1999).
\bibitem{ha} C. Hogan, PRD, 62, 121302 (2000).
\bibitem{hb} C. Hogan, CQG, 18, 4039 (2001).
\bibitem{uv} C. Ungarelli \& A. Vecchio, Phys. Rev. D{\bf 63}, 064030 (2001).
\bibitem{cl} N. J. Cornish \& S. L. Larson, Class. Quant. Grav. {\bf 18}, 3473 (2001).
\bibitem{aet} M. Tinto, J. W. Armstrong \& F. B. Estabrook, Phys. Rev. D{\bf 63}, 021101(R) (2001).
\bibitem{c} N. J. Cornish, Phys. Rev. D{\bf 65}, 022004 (2002).
\end{thebibliography}
\end{document}